\documentclass[aps,twocolumn,amsmath,amssymb,preprintnumbers,superscriptaddress,floatfix]{revtex4}

\usepackage{physics}

\usepackage[utf8]{inputenc}
\usepackage{newtxtext}
\usepackage[upint]{newtxmath}
\usepackage{microtype}
\usepackage{textcomp}
\usepackage{dsfont}
\usepackage{eucal}
\usepackage{bm} 
\usepackage{siunitx}

\usepackage{enumerate}
\usepackage{amsfonts}
\usepackage{amsmath}
\usepackage{amssymb}
\usepackage{color}
\usepackage{soul}

\usepackage{todonotes}
\presetkeys%
    {todonotes}%
    {inline}{}

\usepackage{graphicx}

\usepackage[colorlinks,allcolors=blue]{hyperref}
\usepackage{cleveref}

\renewcommand{\propto}{\sim}

\newcommand{\com}[2]{\left[{#1}\,,\,{#2}\right]}                          
\newcommand{\anticomm}[2]{\left\{{#1}\,,\,{#2}\right\}}                    
\newcommand{\covDer}{\bar{\nabla}}                              

\newcommand{\me}[1]{\mathrm{e}^{#1}}                            
\newcommand{\pd}[2]{\frac{\partial #1}{\partial #2}}            

\newcommand{\cc}[1]{{#1}^*}                                     
\newcommand{\hc}[1]{{#1}^\dagger}                               
\newcommand{\tc}[1]{\tilde{#1}}                                 

\newcommand{\transpose}[1]{\ensuremath{#1}^{\intercal}}

\renewcommand{\v}[1]{\boldsymbol{#1}}                           
\newcommand{\uv}[1]{\v{e}_{#1}}       

\definecolor{DarkRed}{rgb}{0.80,0,0}
\definecolor{Purple}{rgb}{0.55,0,0.55}

\DeclareMathOperator{\re}{Re}
\DeclareMathOperator{\im}{Im}
\DeclareMathOperator{\sgn}{sgn}
\DeclareMathOperator{\diag}{diag}
\DeclareMathOperator{\antidiag}{antidiag}

\newcommand*{\defeq}{\coloneqq}

\renewcommand{\H}[1]{\hat{#1}}
\newcommand{\V}[1]{\check{#1}}

\newcommand{\up}{\uparrow}                                      

\newcommand{\ordOne}{\Psi_1}
\newcommand{\ordTwo}{\Psi_2}
\newcommand{\thouless}{\varepsilon_{\textsc t}}

\newcommand{\ie}{i.e.\ }

\let\epsilon\varepsilon

\begin{document}
\title{Superconducting Vortices in Half-Metals}
\author{Eirik Holm Fyhn}
\affiliation{Center for Quantum Spintronics, Department of Physics, Norwegian \\ University of Science and Technology, NO-7491 Trondheim, Norway}
\author{Jacob Linder}
\affiliation{Center for Quantum Spintronics, Department of Physics, Norwegian \\ University of Science and Technology, NO-7491 Trondheim, Norway}
\date{\today}
\begin{abstract}
  \noindent When the impurity mean free path is short, only spin-polarized Cooper pairs which are non-locally and antisymmetrically correlated in time may exist in a half-metallic ferromagnet. As a consequence, the half-metal acts as an odd-frequency superconducting condensate. 
	We demonstrate both analytically and numerically that quantum vortices can emerge in half-metals despite the complete absence of conventional superconducting correlations.
  Because these metals are conducting in only one spin band, we show that a circulating spin supercurrent accompanies these vortices.
  Moreover, we demonstrate that magnetic disorder at the interfaces with the superconductor influences the position at which the vortices nucleate.
This insight can be used to help determine the effective interfacial misalignment angles for the magnetization in hybrid structures, since the vortex position is experimentally observable via STM-measurements.
  We also give a brief discussion regarding which superconducting order parameter to use for odd-frequency triplet Cooper pairs in the quasiclassical theory.
\end{abstract}
\maketitle
\section{Introduction}
New physical phenomena can emerge at the interface between materials with different quantum order.
One such example is in systems combining ferromagnetism and superconductivity, where it is possible to generate Cooper pairs that are both spin-polarized and correlated non-locally in time.
This has become the basis for the field of superconducting spintronics~\cite{linder2015}, which has as one of its goals to enable new types of devices utilizing spin-polarized supercurrents~\cite{Eschrig2015}.
On a more fundamental level, it is of interest to consider the interplay between different types of spontaneous symmetry breaking in such hybrid structures, since symmetry breaking governs a wide range of physical phenomena, including 
mass differences of elementary particles and phase transitions.

Half-metallic ferromagnets are \SI{100}{\percent} spin-polarized, meaning that only one spin-band is conducting.
Any supercurrent flowing through such a material, as has been observed experimentally~\cite{Keizer2006}, is therefore necessarily spin-polarized.
This makes them especially interesting to study in order to understand how superconductivity adapts to a fully polarized environment.
Much experimental and theoretical work has recently been conducted in order to understand hybrid structures involving superconductors (S) and half-metals (H)~\cite{Keizer2006,Kalcheim2012,singh_prx_15,Pena2004,Eschrig2007,Eschrig2003,eschrigLinder2015,Ouassou2017,MatthiasGeilhufe2009, wu_prb_18, mironov_prb_15, anwar_prb_10, asano_prb_07}.

One hitherto unsolved problem is whether superconducting quantum vortices can form in half-metallic materials. This is an unusual physical situation since the electrons are correlated exclusively non-locally in time, such that the half-metal in fact mimicks a purely odd-frequency~\cite{linder_rmp_19} superconducting state.
Vortices have non-superconducting cores and a phase winding of an integer multiple of $2\pi$ in the superconducting order parameter, leading to circulating supercurrents~\cite{Kwok2016}.
In addition to being interesting from a fundamental physics point of view, understanding the behaviour of vortices is useful on a practical level.
Their motion is a source of non-zero electrical resistance~\cite{Halperin2010}, and recently it has been proposed that vortices can be used as a means for long-range spin transport~\cite{Kim2018}.

\begin{figure}
  \centering
    \includegraphics[width=1.0\linewidth]{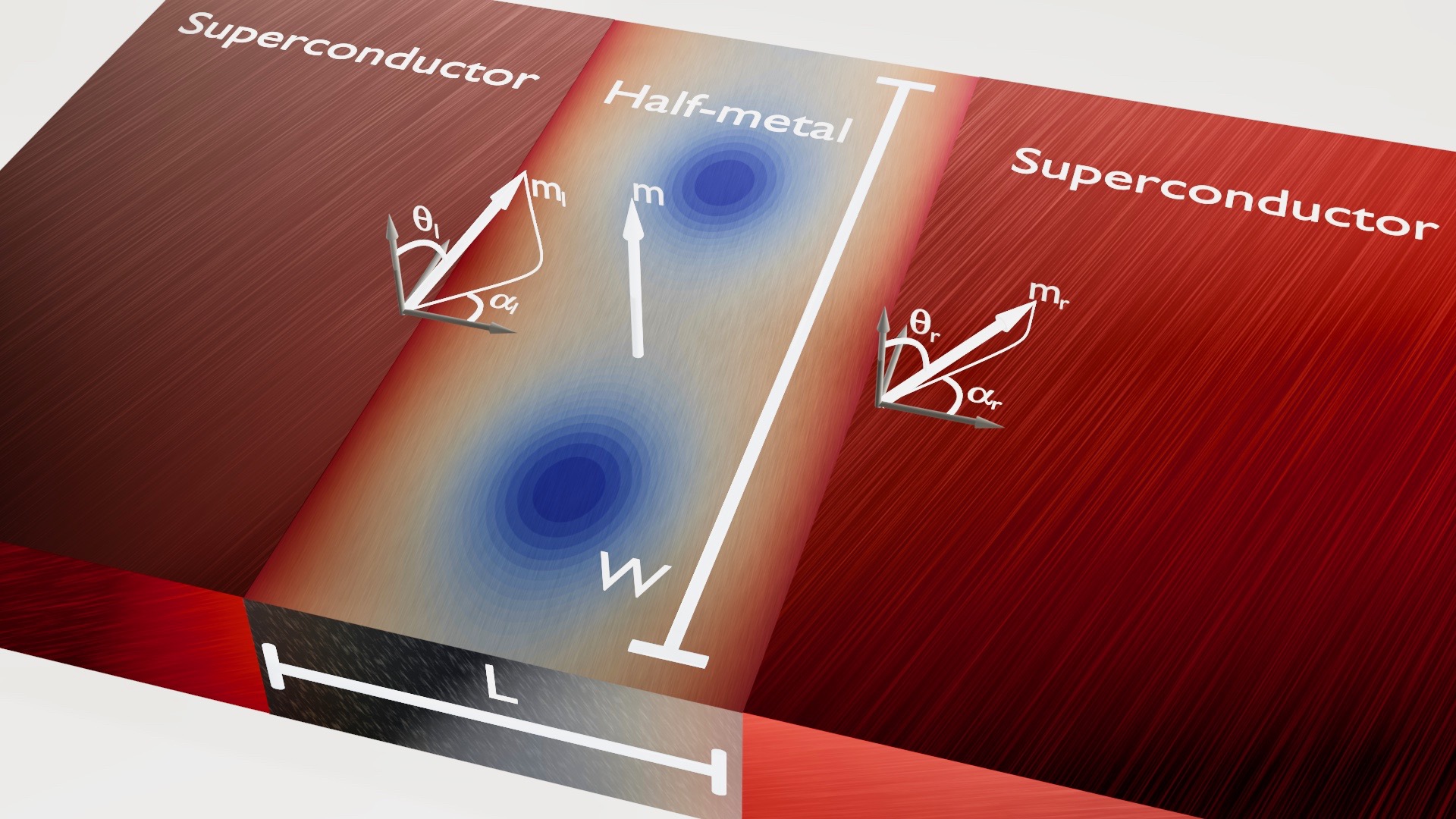}
    \caption{Sketch of a SHS junction. The half-metal (H) has a uniform magnetization direction $\v m$. The interfaces with the left and right superconductor has effective magnetization directions $\v m_l$ and $\v m_r$, respectively. These may come from for instance from artificially inserted thin ferromagnetic layers or from interfacial magnetic disorder.
    $\theta_l$ and $\theta_r$ are the associated polar angles and $\alpha_l$ and $\alpha_r$ are the associated azimuthal angles. $L$ and $W$ are the length and width of the half-metal, respectively.
  The blue regions show the vortex cores where the DOS equals its normal-state value. Away from the cores, the DOS deviates from its normal-state value due to the superconducting proximity effect.
  }
  \label{fig:2DSHMS}
\end{figure}

It is known that vortices can form also inside normal metals that are in the proximity to a superconductor~\cite{Cuevas2007,Bergeret2008a,amundsen_prl_18,Stolyarov2018}.
Cooper pairs can then leak into the normal metal through the process of Andreev reflection~\cite{Pannetier2000}.
This is the key mechanism behind the proximity effect which consists of weak superconductivity observed in a material placed in contact with a superconductor.

The proximity effect in half-metals is more complicated because it requires a mechanism which converts the spinless (singlet) Cooper pairs to spin-polarized (triplet) pairs.
The theorized mechanism to produce such correlations involve spin mixing and spin-flip scattering at the interface~\cite{Eschrig2003}.
Spin mixing introduces triplet correlations at the superconducting side, and spin-flip scattering mediates these correlations to the half-metallic side.

What allows us to investigate SH-heterostructures in the presence of an external magnetic field is the recent derivation of general spin-active boundary conditions for the quasiclassical theory applied to diffusive systems~\cite{eschrigLinder2015,Ouassou2017}.
This means that we can apply the quasiclassical Usadel theory in such a way that the Cooper pair conversion mechanism described above is included.

Here, we apply this theory both analytically and numerically to a two-dimensional SHS-junction as depicted in \cref{fig:2DSHMS} under a constant perpendicular magnetic field.
The constant magnetic field will in general have contributions both from the uniform magnetization in the half-metal and from an uniform applied field.
We find that vortices indeed form in the purely triplet odd-frequency superconducting condensate existing in the half-metallic ferromagnet. 
Their location depends not only on the superconducting phase difference, but also on the effective interfacial magnetization directions characterizing either magnetic disorder or artificially inserted thin ferromagnetic layers \cite{singh_prx_15}.

\section{Methodology}%
In this section we discuss the quasiclassical Usadel theory and how it may be used to analyse the SHS-junction depicted in \cref{fig:2DSHMS}.
We first present the mathematical tools and end with the numerical implementation.
\label{sec:methodology}
\subsection{Quasiclassical theory}
The SHS junction depicted in \cref{fig:2DSHMS} can be treated in the quasiclassical formalism under the assumption that the Fermi wavelength is much shorter than all other relevant length scales.
This assumption is seemingly broken in the half-metal, where the exchange field is so large that the associated energy is not negligible compared to the Fermi energy~\cite{schlottmann2003}.
The redemption comes from the realization that the spin-splitting in such system is so severe that there is effectively no interaction between different spin bands.
We can then continue to use the quasiclassical theory if we, instead of including an exchange field, treat the spin-bands as non-interacting.
If in addition the system is diffusive, meaning that the scattering time is small, the isotropic part of the quasiclassical Green's function dominates and solves the Usadel equation~\cite{Belzig1999,Chandrasekhar2004,Rammer2004,Usadel1970},
\begin{align}
  D\covDer\cdot\left(\V{g}\covDer\V{g}\right) + i\com{\epsilon\H{\rho}_3 + \H{\Delta}}{\V{g}} = 0.
  \label{eq:usadel}
\end{align}
Here, $D$ is a diffusion constant, $\H\rho_3 = \diag(1, 1, -1, -1)$ and $\H{\Delta} = \antidiag(+\Delta,-\Delta,+\Delta^*,-\Delta^*)$ where $\Delta$ is the superconducting gap parameter.
In the superconductors $D$ is a scalar while in the haf-metal it is $\diag(D, 0)$, assuming that the conducting band is spin-up.
The covariant derivative is $\covDer \V g = \nabla \V g - ie\com{\H \rho_3{\v A}}{\V g}$,
where $\v A$ is the vector potential, and
\begin{align}
  \V{g} =
  \begin{pmatrix}
    \H{g}^\textsc{r} & \H{g}^\textsc{k} \\
    0   & \H{g}^\textsc{a}
  \end{pmatrix}
  \label{eq:greenFull}
\end{align}
is the quasiclassical impurity-averaged Green's function.
$\V g$ is normalized such that $\V g \V g = 1$.
We use the convention that when two matrices of different dimensionality is multiplied, the smaller matrix is elevated to the dimensionality of the larger matrix by the tensor product of an identity matrix of the appropriate size.
In equilibrium, the components of the $8 \times 8$ Green's function in \cref{eq:greenFull} are related by the identities $\H{g}^\textsc{k} = \left(\H{g}^\textsc{r} - \H{g}^\textsc{a}\right)\tanh(\varepsilon\beta/2)$ and $\H{g}^\textsc{a} = -\H{\rho}_3 \H{g}^{\textsc{r}\dagger} \H{\rho}_3$, which means that in this case it is sufficient to solve for the retarded component~$\H{g}^\textsc{r}$.

The Usadel equation can be made dimensionless by introducing the Thouless energy, $\thouless \defeq D/L^2$, and measuring length scales relative to $L$ and energies relative to $\thouless$.

In general, the Usadel equation has to be solved together with the Maxwell equation  in a self-consistent manner. However, we are interested here in the case where the width $W$ is smaller than the Josephson penetration depth. In this case one can ignore the screening of the magnetic field by the Josephson currents and the magnetic field is equal to the external one~\cite{Barone1982}.

To simplify the numerical and analytical calculations we assume that the magnetic field is zero inside the superconductors.
This assumption is widely used~\cite{Cuevas2007,Alidoust2015,Bergeret2008a} and has been shown to give good agreement with experimental results~\cite{chiodi2012}.
\citet{belzig1998} found that including the vector potential in the superconductors leads to corrections proportional to $\lambda_s/L$, where $\lambda_s$ is the effective penetration depth in the superconductors and $L$ is the length of the junction.
Assuming $\lambda_s/L$ is small, we neglect the corrections from the vector potential to the Green's function in the superconductors.
We use the vector potential $e\v A = -n\pi y/W \left[\theta(x) - \theta(x - L)\right]\uv x$, where $n = \Phi/\Phi_0$ is the number of flux quanta penetrating the half-metal and $\uv x$ is the unit vector in the $x$-direction.

\subsection{Boundary Conditions}
The quasiclassical formalism is not applicable across boundaries because the associated length scale is too short.
The Usadel equation must therefore be solved in the half-metal and superconductors separately, and the solutions must be connected through boundary conditions.
These can be written
\begin{align}
  \label{eq:bc}
  G_iL_i \uv{n} \cdot (\H g^{\textsc{r}}_i \covDer \H g^{\textsc{r}}_i) = \H I(\H g^{\textsc{r}}_i, \H g^{\textsc{r}}_j),
\end{align}
where $\uv n$ is the outward-pointing normal vector for region $i$, $G_i$ is the bulk conductance of material $i$ and $L_i$ is the length of material $i$ in the direction of $\uv n$.
$\H I(\H g^{\textsc{R}}_i, \H g^{\textsc{R}}_j)$ is the matrix current from material $i$ to material $j$.

The matrix current for general spin-active boundaries between diffusive materials was found in 2015 by \textcite{eschrigLinder2015}.
The expression to second order in spin-mixing angles and transmission probabilities was simplified in 2017 by \textcite{Ouassou2017}, making them easier to implement and more efficient to compute.
This is the expression we will use here, and it reads
\begin{align}
  \label{eq:matCurr}
  \H I &=
  \frac{G_0^i}{2}\com{\H g_i}{F\left(\H g_j\right)} - \frac{iG_{\phi}^i}{2}\com{\H g_i}{\H m_i} + \frac{G_2^i}{8}F\left(\H g_j\right)\H g_i  F\left(\H g_j\right) \notag\\
  &+ \frac{G_{\phi 2}^i}{8}\com{\H g_i}{\H m_i\H g_i\H m_i} + \frac{iG_{\chi}^i}{8}\com{\H g_i}{F\left(\H g_j\right)\H g_i \H m_i + \H m_i \H g_i F\left(\H g_j\right)}\notag\\
  &+ \frac{iG_{\chi}^j}{8}\com{\H g_i}{F\left(\H g_j\H m_j \H g_j - \H m_j\right)},
\end{align}
where, for a half-metallic ferromagnet $F(\H{v}) = \H{v} + \anticomm{\H{v}}{\H{m}} + \H{m} \H{v} \H{m}$ and $\H m_k = \diag(\v m_k \cdot \v \sigma, \v m_k \cdot \cc{\v \sigma})$.
Here $\v \sigma = \transpose{(\sigma_i, \sigma_2, \sigma_3)}$ is the vector of Pauli matrices and $\v m_k$ is a unit vector in the direction of the magnetization experienced by a particle being reflected in material $k$.
Similarly, $\H m = \diag(\v m \cdot \v \sigma, \v m \cdot \cc{\v \sigma})$ where $\v m$ is the unit vector in the direction of the magnetization being felt by a particle which is transmitted.
The interface conductances are~\cite{Ouassou2017}
\begin{align}
  \begin{split}
    G_0^i &= G_q \sum_{n = 1}^N T^i_n,\qquad
    G_{\phi}^i = 2G_q \sum_{n = 1}^N \phi^i_n,\\
    G_{2}^i &= G_q \sum_{n = 1}^N \left(T^i_n\right)^2,\quad
    G_{\chi}^i = G_q \sum_{n = 1}^N T^i_n\phi^i_n,\\
    G_{\phi 2}^i &= 2G_q \sum_{n = 1}^N \left(\phi^i_n\right)^2,
  \end{split}
\end{align}
where $T^i_n$ and $\phi^i_n$ are respectively the transmission probability and spin mixing angle for tunneling channel $n$ from material $i$ to material $j$.
For boundaries interfacing vacuum at $y = \pm W/2$, the matrix current is $\H I = 0$.

\subsection{Ricatti Parametrization}%
\label{sub:ricatti_parametrization}
In the Ricatti parametrization~\cite{schopohl_arxiv_98} of $\H g^{\textsc R}$, the parameter is the $2\times 2$ matrix $\gamma$ and the retarded Green's function is written
\begin{align}
  \H g^{\textsc R} =
  \begin{pmatrix}
    N & 0 \\
    0 & -\tc N
  \end{pmatrix}
  \begin{pmatrix}
    1 + \gamma \tc \gamma & 2\gamma \\
    2\tc\gamma & 1+ \tc\gamma\gamma
  \end{pmatrix},
\end{align}
where $N \defeq \left(1 - \gamma\tc\gamma\right)^{-1}$ and tilde conjugation is $\tc\gamma(\varepsilon) = \cc\gamma(-\varepsilon)$.

There is only one conducting spin band in a half-metal, and as a result $\gamma$ has only one nonzero element,
\begin{equation}
  \gamma_{\textsc{hm}} =
  \begin{pmatrix}
    a & 0\\
    0 & 0
  \end{pmatrix}.
  \label{eq:ricattiHM}
\end{equation}
Substituting this into \cref{eq:usadel} we get that $a$ solves the equation
\begin{align}
\label{eq:usadelRicatti}
  \nabla^2 a + \frac{2\tc a \nabla a \cdot \nabla a}{1 - a\tc a}  = \frac{4(1 + a \tc a)e\v A\cdot (ae\v A + i\nabla a)}{1 - a\tc a} - 2i\varepsilon a,
\end{align}

In \cref{sub:analytics} we will show that the Green's function in the superconductors can be taken to be taken as the bulk Green's function.
Thus, the Ricatti parameter can be written as $\gamma_{\textsc sc} = \antidiag(b, -b)$, where $b$ is a function of $\varepsilon$ and the superconducting gap parameter $\Delta$.
Inserting this and \cref{eq:matCurr,eq:ricattiHM} into \cref{eq:bc} we get
\begin{align}
  \begin{split}
  G_{\textsc{hm}}\uv n \cdot \nabla a = 4G_0^{\textsc{hm}}BCa
  - G_2^{\textsc{hm}}B^2 C^2 a(a\tc a + 3) \\
  + 2i G_{\chi}^{\textsc{sc}}BC^2\sin\theta \left(b\me{-i\alpha} - \tc b \me{i\alpha} a^2\right)
  + 2iG_{\textsc{hm}}a \uv n \cdot \v A e,
  \end{split}
  \label{eq:bcRicatti}
\end{align}
where $B = b\tc b -1$, $C = 1/\left(1 + b\tc b\right)$ and $\theta$ and $\alpha$ are the angles for the magnetization directions on the superconducting side as shown in \cref{fig:2DSHMS}.
The corresponding equations for $\tc a$ and $\uv n \cdot \nabla \tc a$ is found by tilde conjugating \cref{eq:usadelRicatti,eq:bcRicatti}.

From \cref{eq:bcRicatti} it can be seen that in order to have a non-zero solution for $a$ it is necessary that either  $\sin\theta_l\neq 0$ or $\sin\theta_r\neq 0$.
This means that having the effective magnetization angles at the interface  not parallel with the uniform magnetization of the half-metal is necessary for the occurence of superconducting triplet correlations in the half-metal.
This is because the creation of long-range spin-triplet require spin mixing and spin-flip scattering/spin rotation~\cite{Eschrig2003,linder2015}.
When spin-singlet Cooper pairs in a superconductors encounter an interface with effective magnetization direction $\v m_l$, the spin-active boundary will produce spin-triplets with zero spin along $\v m_l$.
If $\v m_l$ is not parallel to the magnetization of the half-metal, $\v m$, then the triplet with zero spin along $\v m_l$ will have a non-zero projection onto the spin-triplet state with spin one along $\v m$.
Hence, it the interfacial magnetization angles are not parallel with $\v m$, the spin-active boundaries will produce equal spin triplets.

\subsection{Observables}
As mentioned initially, a vortex is accompanied by a normal-state density of states and a circulating supercurrent. 
This can be extracted from the quasiclassical Green's function.
In the following it will be useful to write
\begin{equation}
  \H g^{\textsc R} =
  \begin{pmatrix}
    g & f \\
    -\tc f & -\tc g
  \end{pmatrix}.
  \label{eq:gr}
\end{equation}
In the half-metal, $f$ has only one nonzero component, $f_\up$.

\subsubsection{Local Density of States}%
\label{ssub:local_density_of_states}
The local density of states for spin-band $\sigma$ at energy $\varepsilon$ and location $\v r$ can be written
\begin{equation}
  N_\sigma(\varepsilon, \v r) = N_0\real\{g_{\sigma\sigma}(\varepsilon, \v r)\},
  \label{eq:DOS}
\end{equation}
where $N_0$ is the normal state density of state at the Fermi surface.
In the half-metal we can write \cref{eq:DOS} in terms of $a$,
\begin{equation}
  N(\varepsilon, \v r) \defeq N_\up(\varepsilon, \v r) = N_0 \frac{1 + a \tc a}{1 - a \tc a}.
  \label{eq:DOSHM}
\end{equation}

\subsubsection{Supercurrent}

Written in terms of the quasiclassical Green's function, the current density is~\cite{Belzig1999}
\begin{equation}
  \v j = \frac{N_0 eD}{4} \int_{-\infty}^{\infty}\Trace\left(\H\rho_3 \left[\V g \covDer\V g\right]^{\textsc K}\right) \dd{\varepsilon}.
  \label{eq:currentGen}
\end{equation}
Inserting \cref{eq:gr}, using the relations $\H{g}^\textsc{a} = -\H{\rho}_3 \H{g}^{\textsc{r}\dagger} \H{\rho}_3$, $\H{g}^\textsc{k} = \left(\H{g}^\textsc{r} - \H{g}^\textsc{a}\right)\tanh(\varepsilon\beta/2)$,
\cref{eq:currentGen} can be rewritten
\begin{align}
  \begin{split}
  \v j = \frac{N_0 eD}{2} \int_{-\infty}^{\infty}\tanh\left(\frac{\beta\varepsilon}{2}\right)\Trace\Bigl(\re\left[\hc{\tc f}\nabla \hc f - f \nabla\tc f\right] \\
  + 2e\v A \im\left[f\tc f - \hc{\tc f}\hc f\right]\Bigr) \dd{\varepsilon}.
  \end{split}
\end{align}

The spin current can be found by multiplying the matrix in the integrand of \cref{eq:currentGen} by the Pauli matrix corresponding to the appropriate spin direction before taking the trace.
For a half-metal magnetized in the $z$-direction, the $z$-component of the spin supercurrent polarization is proportional to the electric current while the remaining spin current components vanish.

\subsubsection{Cooper Pair Correlation Function}
The study of vortices in diffusive half-metals naturally raises the question of how to define the superconducting order parameter.
In a normal superconductor, the order parameter is $\expval{\psi_\uparrow(\v r, 0)\psi_{\downarrow}(\v r, 0)}$ where $\psi_{\sigma}(\v r, t)$ is the field operator which destroys an electron with spin $\sigma$ at position $\v r$ and time $t$.
The same order parameter is used in a normal metal, but the analogous quantity for the half-metal, $\expval{\psi_\uparrow(\v r, 0)\psi_{\uparrow}(\v r, 0)}$ is always zero because of the Pauli exclusion principle.
That is, the Cooper pair correlation function in a diffusive half-metal must vanish at equal times and is thus temporally non-local~\cite{linder_rmp_19}.

One approach of defining an order parameter in odd-frequency superconducting condensates, which is often used in the Bogolioubov-de Gennes formalism~\cite{halterman_prl_07}, is to keep the relative time coordinate $t$ finite between the field operators, that is
\begin{multline}
  \ordOne(\v r, t) \defeq \expval{\psi_\uparrow(\v r, t)\psi_{\uparrow}(\v r, 0)} \\
  = \frac{-iN_0}{2}\int_{-\infty}^{\infty}f_{\uparrow}(\v r, \varepsilon)\tanh(\varepsilon\beta/2)\sin(\varepsilon t)\dd{\varepsilon}.
\end{multline}
Another frequently used strategy~\cite{abrahams_prb_95} is to make the order parameter even in time by differentiation.
This yields
\begin{align}
  \ordTwo(\v r) \defeq \left.\pd{\ordOne}{t}\right|_{t=0}
  = \frac{-iN_0}{2}\int_{-\infty}^{\infty}\varepsilon f_{\uparrow}(\v r, \varepsilon)\tanh(\varepsilon\beta/2)\dd{\varepsilon}.
\end{align}
Below, we shall compare these two possible choices for order parameter describing the odd-frequency superconducting condensate to see which of them that correctly captures the vortex behavior.

\subsection{Numerics}
The Usadel equation was solved numerically using a finite element scheme.
See for instance~\cite{Amundsen2016} to see how to set up solve the nonlinear Usadel equations in a finite element scheme by the use of the Newton-Rhapson method.
The program was written in Julia~\cite{Bezanson2017}, we used quadratic quadrilateral elements and JuAFEM.jl~\cite{Carlsson2019} was used to iterate through the cells.
Gauss-Legandre quadrature rules of fourth order was used to integrate through the cells and Romberg integration was used to integrate over energy.
See for instance~\cite{sauer2013}.
Finally, forward-mode automatic differentiation~\cite{RevelsLubinPapamarkou2016} was used to calculate the Jacobian.

\section{Results and Discussion}%
Here we present first an analytical solution of the Usadel equation in the weak proximity effect regime, then we show numerically that the findings is present also in the full proximity effect regime.
Dimensionless quantities are used in the analytics with length being measured relative to the length of the half-metal, $L$, and energies being measures relative to the Thouless energy $\thouless = D/L^2$, where $D$ is the diffusion constant in the half-metal.
\subsection{Analytics}%
\label{sub:analytics}
In order to justify \cref{eq:bcRicatti} we will show that it suffices to use the bulk solution
\begin{equation}
  \H g_{\textsc{bcs}} = \left[\frac{\theta\left(\varepsilon^2 - \abs{\Delta}^2\right)}{\sqrt{\varepsilon^2 - \abs{\Delta}^2}}\sgn(\varepsilon) - \frac{\theta\left(\abs{\Delta}^2 - \varepsilon^2\right)}{\sqrt{\abs{\Delta}^2 - \varepsilon^2}}i \right]\left(\varepsilon\H\rho_3 + \H\Delta\right),
  \label{eq:bulkBCS-hm}
\end{equation}
in the superconductors when a certain condition is fulfilled. Let $\lambda$ (to be defined quantitatively below) be the length-scale over which the Green function recovers its bulk value in the superconductor. The criterion for neglecting the inverse proximity effect in the superconductors is then that the normal-state conductance of the superconductors for a sample of length $\lambda$ is much larger than the interface conductance and that the length of each superconductor is not small compared to $\lambda$. We now proceed to prove this.

First, let
\begin{equation}
  \H g = \H g_{\textsc{bcs}} + \delta \H g
\end{equation}
be the solution of the dimensionfull Usadel equation,
\begin{align}
  D_{\textsc{sc}}\nabla\cdot\left(\H{g}\nabla\H{g}\right) + i\com{\epsilon\H{\rho}_3 + \H{\Delta}}{\H{g}} = 0
  \label{eq:usadel-hm}
\end{align}
in the superconductor at $x<0$.
This gives an equation for $\delta \H g$,
\begin{align}
  D_{\textsc{sc}}\nabla\cdot\left(\left[\H g_{\textsc{bcs}} + \delta \H g\right]\nabla\delta\H{g}\right) + i\com{\epsilon\H{\rho}_3 + \H{\Delta}}{\delta\H{g}} = 0,
  \label{eq:dg}
\end{align}
where we have used that $\H g_{\textsc{bcs}}$ solves the \cref{eq:usadel-hm} for a bulk superconductor and assumed that the variations of the gap parameter from the bulk value is negligible.
Next, assume the inverse proximity effect to be weak, such that $\delta \H g \ll \H g_{\textsc{bcs}}$.
Using that $\H g_{\textsc{bcs}}\H g_{\textsc{bcs}}=1$, this yields
\begin{align}
  D_{\textsc{sc}}\nabla^2\delta\H{g} + i\H g_{\textsc{bcs}}\com{\epsilon\H{\rho}_3 + \H{\Delta}}{\delta\H{g}} = 0.
  \label{eq:dg2}
\end{align}
$\H g_{\textsc{bcs}} + \delta \H g$ must also satisfy the normalization condition $\H g^2 = 1$, so
\begin{equation}
  \left(\H g_{\textsc{bcs}} + \delta \H g\right)^2 = 1 \implies \anticomm{\H g_{\textsc{bcs}}}{\delta\H g} = 0.
\end{equation}
Hence, using that $\com{\epsilon\H{\rho}_3 + \H{\Delta}}{\H g_{\textsc{bcs}}} = 0$,
\begin{align}
  \H g_{\textsc{bcs}}\com{\epsilon\H{\rho}_3 + \H{\Delta}}{\delta\H{g}} &=
  (\epsilon\H{\rho}_3 + \H{\Delta})\H g_{\textsc{bcs}}\delta\H{g} 
  + \delta\H{g}(\epsilon\H{\rho}_3 + \H{\Delta})\H g_{\textsc{bcs}} \notag \\
  &= \anticomm{\delta\H g}{(\epsilon\H{\rho}_3 + \H{\Delta})\H g_{\textsc{bcs}}}.
\end{align}
Finally, from
\begin{equation}
  \left(\epsilon\H{\rho}_3 + \H{\Delta}\right)^2 = \varepsilon^2 - \Delta^2
\end{equation}
we get that $\delta\H g$ is an eigenfunction of the Laplacian,
\begin{equation}
  \nabla^2 \delta\H g = \lambda^{-2} \delta \H g
\end{equation}
where
\begin{align}
  \lambda^{-2} = -\frac{2i}{D_{\textsc{sc}}}\Biggl[\sgn(\varepsilon)\sqrt{\varepsilon^2 - \abs{\Delta}^2}\theta\left(\varepsilon^2 - \abs{\Delta}^2\right) \notag
  \\+ i\sqrt{\abs{\Delta}^2 - \varepsilon^2}\theta\left(\abs{\Delta}^2 - \varepsilon^2\right)\Biggr].
  \label{eq:lambda}
\end{align}
We can choose the sign of $\lambda$ to be such that $\real(\lambda) > 0$.

Using the boundary condition
\begin{equation}
  \nabla \delta\H g\bigr\rvert_{\v r \in \Omega} = 0,
\end{equation}
where $\Omega$ is the boundary not interfacing the half metal, we get
\begin{equation}
  \delta \H g(\varepsilon, x, y) = C\left[ \me{-\abs{x}/\lambda} + \me{\left(\abs{x}-2L_{\textsc{sc}}\right)/\lambda}\right],
  \label{eq:deltag_sol}
\end{equation}
where $C$ is some a function of $y$ and $\varepsilon$ to be determined by the final boundary condition.
If the matrix current across this boundary is $\H I_{\textsc{sc}}$, then
\begin{equation}
  C = \frac{\H g_{\textsc{bcs}}\H I_{\textsc{sc}}}{\left(1 - \me{-2L_{\textsc{sc}}/\lambda}\right)G_{\textsc{sc}}L_{\textsc{sc}}/\lambda}.
\end{equation}
From \cref{eq:deltag_sol} we see that $\real (\lambda)$ can be interpreted as the penetration depth of $\delta g$.
Note that $\real(\lambda)$ is bounded by including the effect of inelastic scattering, which is done by the substitution $\varepsilon \to \varepsilon + i\delta$ for some positive scattering rate $\delta$~\cite{Dynes1984}.
This ensures that $1/\left(1 - \me{-2L_{\textsc{sc}}/\lambda}\right)$ remains finite as $\epsilon \to \Delta$.
From the definition of $\lambda$ in \cref{eq:lambda} we see that $\real (\lambda) = \lambda$ when $\varepsilon < \abs{\Delta}$ and $\real (\lambda ) = \abs{\lambda}/\sqrt{2}$ otherwise.

$G_{\textsc{sc}}$ is the conductance over the whole length $L_{\textsc{sc}}$ of the superconductor and is therefore proportional to $1/L_\textsc{sc}$.
Hence, $G_{\textsc{sc}}L_{\textsc{sc}}/\real(\lambda)$ is the normal-state conductance of a superconductor of length $\real(\lambda)$. 
From the definition of $\H I$ in \cref{eq:matCurr} we see that $C$, and therefore $\delta g$, becomes negligble when
\begin{equation}
  \frac{\max\left(G_0^{\textsc{sc}}, G_{\phi}^{\textsc{sc}}, G_2^{\textsc{sc}}, G_{\phi 2}^{\textsc{sc}}, G_{\chi}^{\textsc{sc}}, G_{\chi}^{\textsc{hm}}\right)}{G_{\textsc{sc}}L_{\textsc{sc}}/\real(\lambda)}  \ll 1,
  \label{eq:inverseCriterion}
\end{equation}
provided that the length of the superconductor $L_\textsc{sc}$ is not small compared to the maximal penetration depth, $\max[\real(\lambda)]$.
A similar calculation shows that we can use $\H g_{\textsc{bcs}}$ also in the superconductor at $x > L$.

Taking the superconducting coherence length $\xi$ as a measure of the inverse proximity effect penetration depth $\real(\lambda)$, we see that the criterion~\cref{eq:inverseCriterion} is indeed experimentally feasible. The equation is fulfilled for a low-transparency interface and for a superconductor that is larger than the coherence length.

Next we turn to the solution of the Usadel equation in the half-metal, \cref{eq:usadelRicatti}, together with the boundary condition~\eqref{eq:bcRicatti}.
In order to solve these equations we must make some simplifying assumptions.
If we assume the proximity to be weak, we can keep only terms which are linear in $a$ and $\tc a$ and their gradients.
In this case the dimensionless Usadel equation~\eqref{eq:usadelRicatti} decouples:
\begin{subequations}
  \label{eq:usadelLinearized}
\begin{align}
  \nabla^2 a = 4\v A\cdot (a\v A + i\nabla a) - 2i\varepsilon a,
  \label{eq:usadelLinearized_a}\\
  \nabla^2 \tc a = 4\v A\cdot (\tc a\v A - i\nabla \tc a) - 2i\varepsilon\tc a.
\end{align}
\end{subequations}
and so does the boundary conditions,
\begin{subequations}
\begin{align}
  \uv n \cdot \nabla a = \left\{4\frac{G_0^{\textsc{hm}}}{G^{\textsc{hm}}}BC
  - 3\frac{G_2^{\textsc{hm}}}{G^{\textsc{hm}}}B^2 C^2 + 2i \uv n \cdot \v A e\right\} a \notag \\
  + 2i \frac{G_{\chi}^{\textsc{sc}}}{G^{\textsc{hm}}}BC^2 \abs{b}\sin\theta \me{i(\phi - \alpha)},
 \\
  \uv n \cdot \nabla \tc a = \left\{4\frac{G_0^{\textsc{hm}}}{G^{\textsc{hm}}}BC
  - 3\frac{G_2^{\textsc{hm}}}{G^{\textsc{hm}}}B^2 C^2 - 2i \uv n \cdot \v A e\right\} \tc a \notag \\
  - 2i \frac{G_{\chi}^{\textsc{sc}}}{G^{\textsc{hm}}}BC^2 \lvert\tc b\rvert\sin\theta \me{-i(\phi - \alpha)}.
\end{align}
\label{eq:bcSpinLinearized}
\end{subequations}

\Cref{eq:usadelLinearized} can be further simplified in the so-called wide junction limit, where $n/W \ll 1$.
If $\v A = 0$, the solution of \cref{eq:usadelLinearized} is constant in the $y$-direction.
Assuming this is approximately true also for small $\v A$, we neglect the term $\partial_y^2a$.
\Cref{eq:usadelLinearized} can now be solved exactly, as it is a second order ordinary differential equation with constant coefficients.
The solution of \cref{eq:usadelLinearized_a} is
\begin{equation}
  a = C_1 \me{(u + k)x} + C_2 \me{(u - k)x},
\end{equation}
where $u = - 2\pi i ny/W$, $k = \sqrt{-2i\varepsilon}$ and $C_1$ and $C_2$ are independent of $x$.

Determining $C_1$ and $C_2$ requires the boundary conditions, which can be written
\begin{subequations}
  \begin{align}
    \left. \pdv{a}{x} \right|_{x = 0} &= -c\sin\theta_l \me{i(\phi_l - \alpha_l)} - \left(d - u\right)a,
      \\
      \left. \pdv{a}{x} \right|_{x = 1} &= c\sin\theta_r \me{i(\phi_r - \alpha_r)} + \left(d + u\right)a,
  \end{align}
\end{subequations}
where
\begin{align}
  c = 2i \frac{G_{\chi}^{\textsc{sc}}}{G^{\textsc{hm}}}BC^2 \abs{b},
  \qquad
  d = 4\frac{G_0^{\textsc{hm}}}{G^{\textsc{hm}}}BC
  - 3\frac{G_2^{\textsc{hm}}}{G^{\textsc{hm}}}B^2 C^2.
\end{align}
After some algebra, we find that the solution can be written
\begin{widetext}
\begin{align}
  \begin{split}
    a = \frac{c\me{i(\phi_l - \alpha_l) + ux}}{(k - d)^2\me{k} - (k + d)^2\me{-k}}
    \Bigl\{(k - d)\left(\sin\theta_l\me{-k(1-x)} + \sin\theta_r\me{i\delta\phi - u} \me{-k x}\right) 
      +(k + d)\left(\sin\theta_l\me{k(1-x)} + \sin\theta_r\me{i\delta\phi - u}\me{k x}\right) \Bigr\},
  \end{split}
  \label{eq:linSol}
\end{align}
\end{widetext}
where
\begin{equation}
  \delta\phi = \phi_r - \alpha_r - \phi_l + \alpha_l.
\end{equation}
Note that the wide junction approximation is not applicable at small energies.
This is because the solution will be slowly varying in the $x$-direction and therefore $\partial_y^2 a$ is no longer negligible compared to $\partial_x^2 a$.

When $\sin\theta_l = \sin\theta_r$, $a$ vanishes at $x = 1/2$ and
\begin{equation}
  \frac y W =\frac 1 n \left(\frac 1 2 + N - \frac{\phi_r - \alpha_r - \phi_l + \alpha_l}{2\pi}\right),
  \label{eq:vortPosY}
\end{equation}
where $N$ is any integer.
This means that $f_{\up}$ and hence also the order parameters $\ordOne$ and $\ordTwo$ vanish at these points.
From \cref{eq:DOSHM} we see that the density of states is equal to the normal state density of states at these points, indicating that these are indeed vortices.
By Taylor expanding $a$ to first order around a root located at $(1/2, \tilde y)$ we find
\begin{equation}
  a \propto B_1\cos(\theta + \alpha_1) + iB_2\cos(\theta+\alpha_2),
\end{equation}
where $x-1/2 \propto \cos\theta$ and $ y-\tilde y \propto \sin\theta$, $B_1^2 = 5\abs{k}^2/4 - \abs{k}d + 2d^2$, $B_2^2 = \abs{k}^2/4 + d^2$, $\alpha_1 = \tan^{-1}[(\abs{k}/2 + d)/(\abs{k} - d)]$ and $\alpha_2 = \tan^{-1}(\abs{k}/2d)$.
Hence, these roots have a phase winding of $2\pi$, as is characteristic for vortices.
These approximately $n$ roots are the only ones for $\ordTwo$, but for $\ordOne$ there are relative times $t$ for which additional roots exist.
Since each vortex is associated with a quantum of magnetic flux, $\Phi_0$, there should be at most $n$ vortices when the flux is $n\Phi_0$.
This suggests that $\ordOne$ is less suited for finding vortices than $\ordTwo$ if we identify vortices by the roots of the order parameter.
Using $\ordTwo$ suggests that when $\sin\theta_l = \sin\theta_r$ and the magnetic flux is $n\Phi_0$, there will in the wide-junction limit be $n$ vortices whose location is determined by the difference in the superconducting phases and the magnetization angles.

The situation is more complicated when $\sin\theta_l \neq \sin\theta_r$.
In this case the roots of \cref{eq:linSol} depend on $\varepsilon$, and we will leave the discussion for how this affects the order parameter to the numerical investigation.
However, some insight can still be had from the analysis.
Scaling $\sin\theta$ in the boundary condition~\eqref{eq:bcSpinLinearized} is equivalent to scaling the conductance $G_{\chi}^{\textsc{sc}}$.
That is, if $\sin\theta_r < \sin\theta_l$, the proximity effect should be weaker at the right side, meaning that the vortices should be pushed to the right.
This is indeed what we find numerically.

\subsection{Numerics}%
\label{sec:numerics}
We now proceed to show numerical results in the full (non-linear) proximity effect regime.
We have set the parameters $\abs{\Delta} = 4\thouless$, $G_{\textsc{hm}} = 3 G_0^{\textsc{hm}}$, $G_{\chi}^{\textsc{sc}} = 0.01G_0^{\textsc{hm}}$, $G_{2}^{\textsc{hm}} = 0.002G_0^{\textsc{hm}}$ and $\phi_l = \alpha_l = 0$ common for all the numerical calculations.
We obtain qualitatively similar results for other choices of the conductance parameters $G_i^\textsc{hm/sc}$.
We include the effect of inelastic scattering by doing the substitution $\varepsilon \to \varepsilon + i\delta$ where $\delta = 0.001\abs{\Delta}$ in order to avoid the divergence of $\H g_{\textsc{bcs}}$ at $\varepsilon = \abs{\Delta}$~\cite{Dynes1984}.

\subsubsection{Local Density of States}
\label{ssub:local_density_of_states_res}
\begin{figure}
  \centering
  \includegraphics[width=1.0\linewidth]{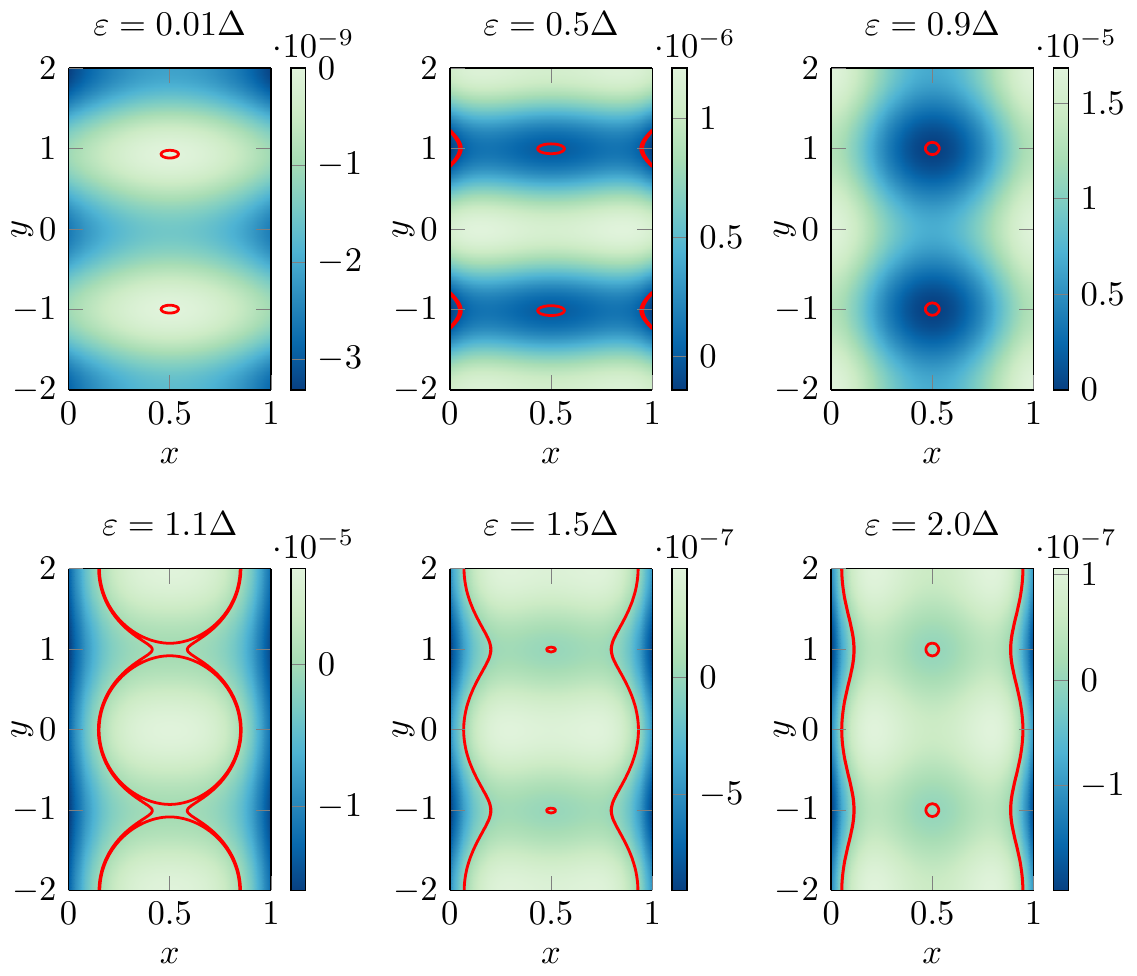}
    \caption{The normalized difference between the local density of states and the normal-state density of states, $(N - N_0)/N_0$, for various energies.
    Red contour lines are added at $\pm 0.01\times S$, where $S$ is the number at the top of the respective colorbars. Here $n=2$, $\phi_r=\alpha_r=0$ and $\sin\theta_r = 1$.}
    \label{fig:symDos}
\end{figure}

In the symmetric case ($\sin\theta_r = 1.0$), we suspect from the analysis above that for all energies, the local density of states is equal to that of the normal state in $n$ points along the line $x = 0.5$, where $n$ is the number of magnetic flux quanta.
This is also what we find numerically, as shown in \cref{fig:symDos}.
\Cref{fig:symDos} shows the local density of states at various energies for the symmetric case with $n=2$ with red lines close to where it is equal to the normal state density of states.
We see that there are indeed two locations where the difference between the local density of states and the normal state density of states vanish for all energies, and that the locations of these points are exactly those predicted by the analysis.
\begin{figure}
  \centering
  \includegraphics[width=1.0\linewidth]{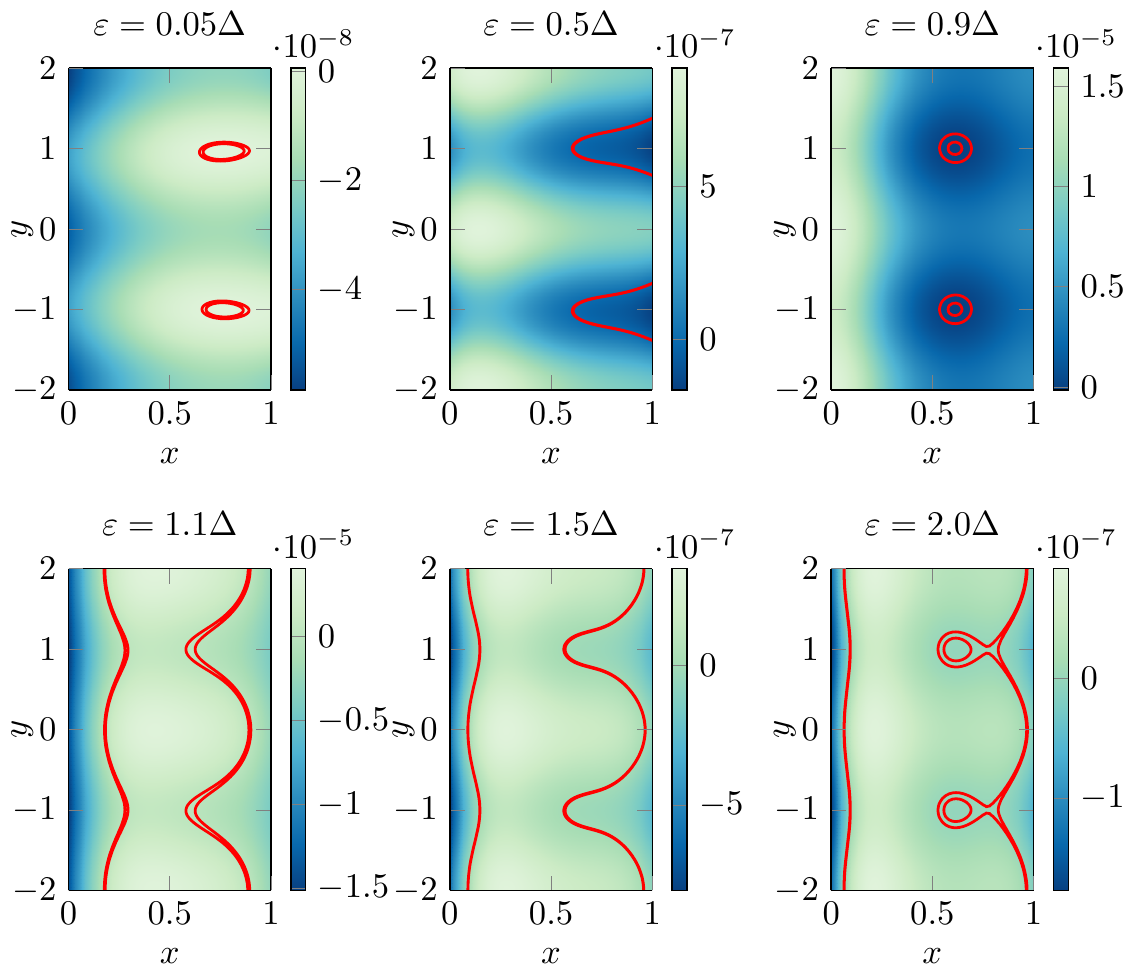}
    \caption{The normalized difference between the local density of states and the normal-state density of states, $(N - N_0)/N_0$, for various energies.
    Red contour lines are added at $\pm 0.01\times S$, where $S$ is the number at the top of the respective colorbars. Here $n=2$, $\phi_r=\alpha_r=0$ and $\sin\theta_r = 0.5$.}
    \label{fig:nonSymDos}
\end{figure}

The situation is slightly different for the asymmetric case ($\sin\theta_r \neq 1$), as can be seen in \cref{fig:nonSymDos}.
There is no longer a single point where $N = N_0$ for all energies.
Instead there are points where $N$ is equal to or almost equal to $N_0$ for all energies, as can be seen in \cref{fig:nonSymDos}.
In \cref{ssub:supercurrent,ssub:cooper_pair_correlation_function} we will se that these points also are associated with vortices.
Qualitatively, the points where $N$ stays close to $N_0$ are different in the asymmetric case.
In the symmetric case the points at $x = 1/2$ are mostly isolated, but in the asymmetric case the point is part of a line where $N = N_0$ which stretches towards the side there $\sin\theta$ is smaller.
For $n=2$, as can be seen in \cref{fig:nonSymDos}, the location where $N$ is close to or equal to $N_0$ for alle nergies occurs at $y = \pm W/4$, which is also the $y$-values where the vortices are in the symmetric case.

Note that the energy-dependence on the position where $N = N_0$ is a not unique for the special case of a SHS-junction.
The same phenomenon occurs in normal SNS-junctions if the conductances at the interfaces are unequal.
As mentioned above, changing $\sin\theta$ is the same as changing the conductance $G_{\chi}^{\textsc{sc}}$.

\subsubsection{Supercurrent}
\label{ssub:supercurrent}
\begin{figure}
  \centering
  \includegraphics[width=1.0\linewidth]{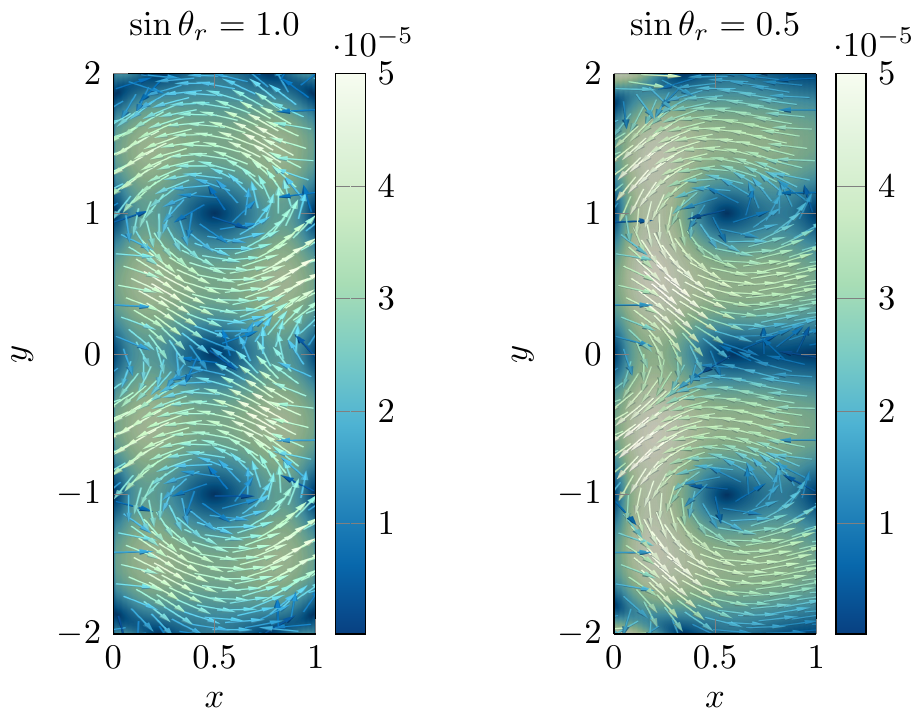}
    \caption{Amplitude and direction of the fully spin-polarized supercurrent $\v j$ for the symmetric case ($\sin\theta_r = 1$) and the asymmetric case ($\sin\theta_r \neq 1$). Here $n=2$ and $\phi_r=\alpha_r=0$. The values are given for the dimensionless supercurrent $\v j \times (L/N_0 e D \thouless)$.}
    \label{fig:curr}
\end{figure}

\Cref{fig:curr} shows the current amplitude and direction for the same two cases as was discussed in \cref{ssub:local_density_of_states_res}.
In both cases there are exactly two points where the supercurrents vanish and where the supercurrent circles around.
This indicates the existence of two superconducting vortices, which is in accordance with the analysis, local density of states and the fact that the system is experiencing two quanta of magnetic flux. We underline that the supercurrents accompanying the vortices induced in the half-metal are fully spin-polarized and carried by triplet Cooper pairs. This is different from the non-polarized charge supercurrents circulating vortices in previously studied hybrid structures \cite{Cuevas2007, Bergeret2008a, amundsen_prl_18}.

The $y$-value of the points with circulating supercurrents are the same as what is expected by the analysis, given by \cref{eq:vortPosY}.
In the symmetric case these points are midway between the superconductors, \ie~at $x=L/2$, while in the asymmetric case they are moved slightly toward the side where $\sin\theta$ is smaller.
Hence, the vortex locations as given by the supercurrents agrees with the analysis as well as the results from the local density of states.

One feature of the asymmetric case worth noting is that the supercurrent is suppressed to the right of the vortex, which is towards the side where $\sin\theta$ is smaller.
This is in agreement with the fact that this region had a local density of states which was closer to the normal state value, as shown in \cref{ssub:local_density_of_states_res}.

\subsubsection{Cooper Pair Correlation Function}
\label{ssub:cooper_pair_correlation_function}
Investigating vortices in odd-frequency superconductors gives rise to the problem of choosing what order parameter to use.
The Cooper pair correlation function which is normally used is not applicable as it is identically equal to zero.
The two most obvious remaining choices are
\begin{equation}
  \ordOne(\v r, t) \defeq \expval{\psi_\uparrow(\v r, t)\psi_\uparrow(\v r, 0)}
  \label{eq:ordParam1_hm2}
\end{equation}
and
\begin{equation}
  \ordTwo(\v r) \defeq \left.\pd{\expval{\psi_\uparrow(\v r, t)\psi_\uparrow(\v r, 0)}}{t}\right|_{t=0},
    \label{eq:ordParam2_hm2}
\end{equation}
as mentioned above.
From looking at the local density of states and the supercurrent we already know that we expect vortices and that their location should have $y$-value given by \cref{eq:vortPosY} and be at $x=L/2$ in the symmetric case and closer to the side where $\sin\theta$ is smaller in the asymmetric case.
Comparing the position of the roots of $\ordOne$ and $\ordTwo$ with the position of the vortices as predicted by the local density of states and the supercurrent can be used to give an identification of how good the order parameters are as means to find vortices.
\begin{figure}
  \centering
  \includegraphics[width=1.0\linewidth]{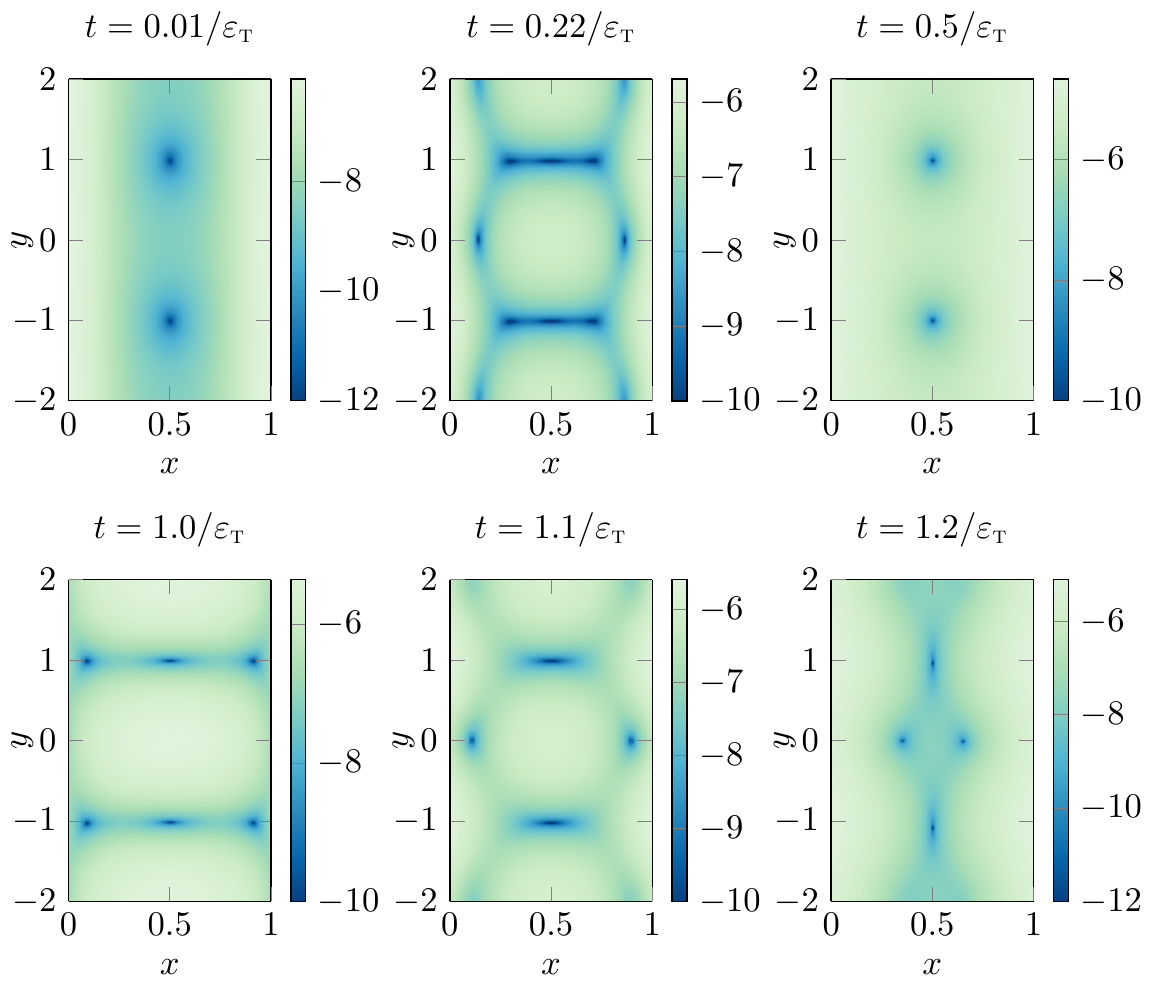}
  \caption{Plots of $\log(\abs{\ordOne(\v r, t)}\times 2/N_0\thouless)$ for various values of $t$ in the symmetric case ($\sin\theta_r=1$) with $n=2$ and $\phi_r=\alpha_r=0$.}
  \label{fig:symTPair}
\end{figure}
\begin{figure}
  \centering
  \includegraphics[width=1.0\linewidth]{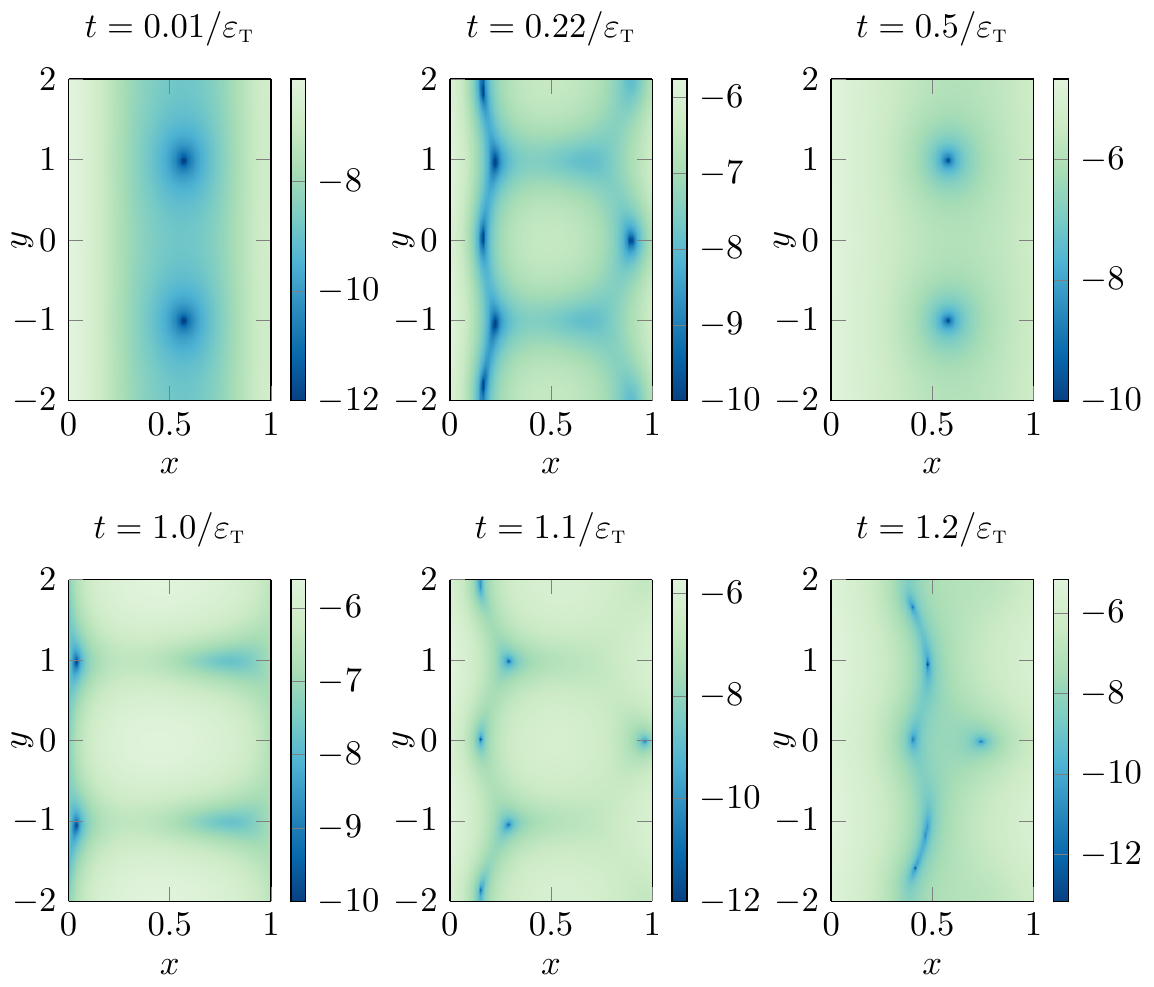}
  \caption{Plots of $\log(\abs{\ordOne(\v r, t)}\times 2/N_0\thouless)$ for various values of $t$ in the symmetric case ($\sin\theta_r=1$) with $n=2$ and $\phi_r=\alpha_r=0$.}
  \label{fig:nonSymTPair}
\end{figure}

\Cref{fig:symTPair,fig:nonSymTPair} shows $\ordOne$ with various values of $t$ for the two same cases which was shown in \cref{ssub:local_density_of_states_res,ssub:supercurrent}.
\Cref{fig:symTPair} shows the symmetric case with $n=2$, $\phi_r=\alpha_r=0$ and $\sin\theta_r=1$, while \cref{fig:nonSymTPair} shows the asymmetric case with $n=2$, $\phi_r=\alpha_r=0$ and $\sin\theta_r=0.5$.
$\ordTwo(\v r)$ looks identical to $\ordOne(\v r, 0.01/\thouless)$, which is shown in the figures, but is a factor $100$ larger in magnitude.
$\ordTwo$ is therefore not shown.
In both cases, there are exactly two roots in $\ordTwo$ and $\ordOne$ for small $t$, and the position are the same as those given for the vortices by the local density of states and supercurrent.
Around these two roots are a phase winding of $2\pi$.

For larger values of $t$, $\ordOne(\v r, t)$ seems to be less suited for finding vortices.
In the symmetric case there are additional roots which appear.
These additional roots also has a corresponding phase winding of $2\pi$, but does not correspond to vortices when compared to the local density of states or supercurrent.
The situation is even worse for large $t$ in the asymmetric case.
In addition to having extra roots, the original roots which correspond to vortices are either moved or not present.
This is also the case for other choices of $n$, $\phi_r$, $\alpha_r$ and $\sin\theta_r$.

Thus, we conclude that using $\ordTwo$ seems best suited as order parameter for the numerical investigation of quantum vortices in an purely odd-frequency superconducting condensate.
Alternatively, one could use $\ordOne$ with $\thouless t \ll 1$ which will give the same result since $\ordOne \sim \ordTwo t$ as $t\to 0$.

\Cref{fig:sym,fig:nonSym} shows the phase of $\ordTwo$, in addition to the supercurrent and amplitude of $\ordTwo$, for a magnetic flux of $4\Phi_0$ and with $\sin\theta_r = 1$, $\phi_r = \alpha_r = 0$ and $\sin\theta_r = 0.5$, $\phi_r - \alpha_r = \pi/2$, respectively.
Note that there is indeed a phase winding of $2\pi$ around the vortices, as mentioned above.

\subsubsection{Vortex Position}
\begin{figure}
  \centering
  \includegraphics[width=1.0\linewidth]{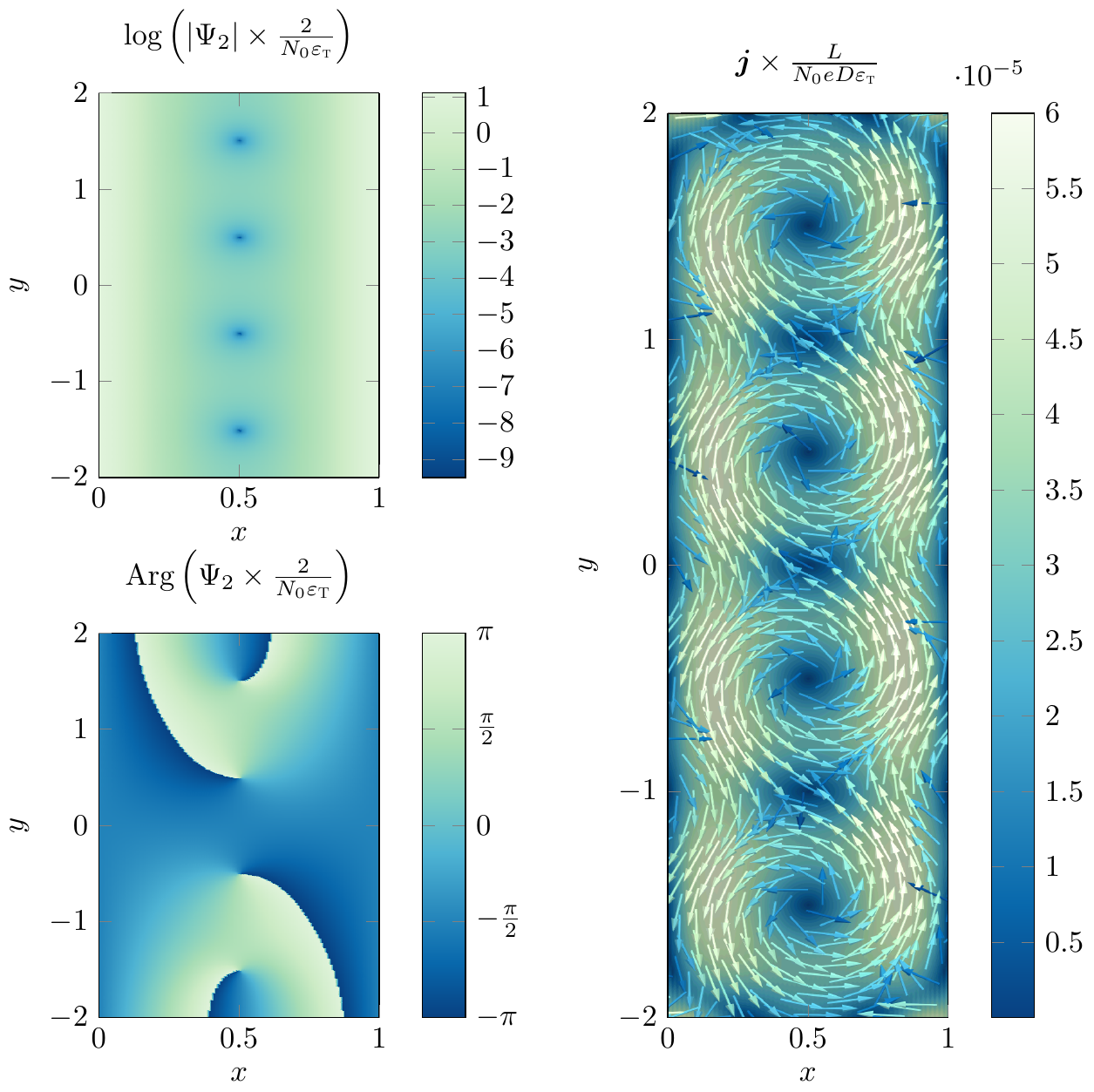}
    \caption{Amplitude and phase of the order parameter $\ordTwo$ and amplitude and direction of the fully spin-polarized supercurrent $\v j$. Here $n=4$, $\phi_r=\alpha_r=0$ and $\sin\theta_r = 1$.}
  \label{fig:sym}
\end{figure}
\begin{figure}
  \centering
  \includegraphics[width=1.0\linewidth]{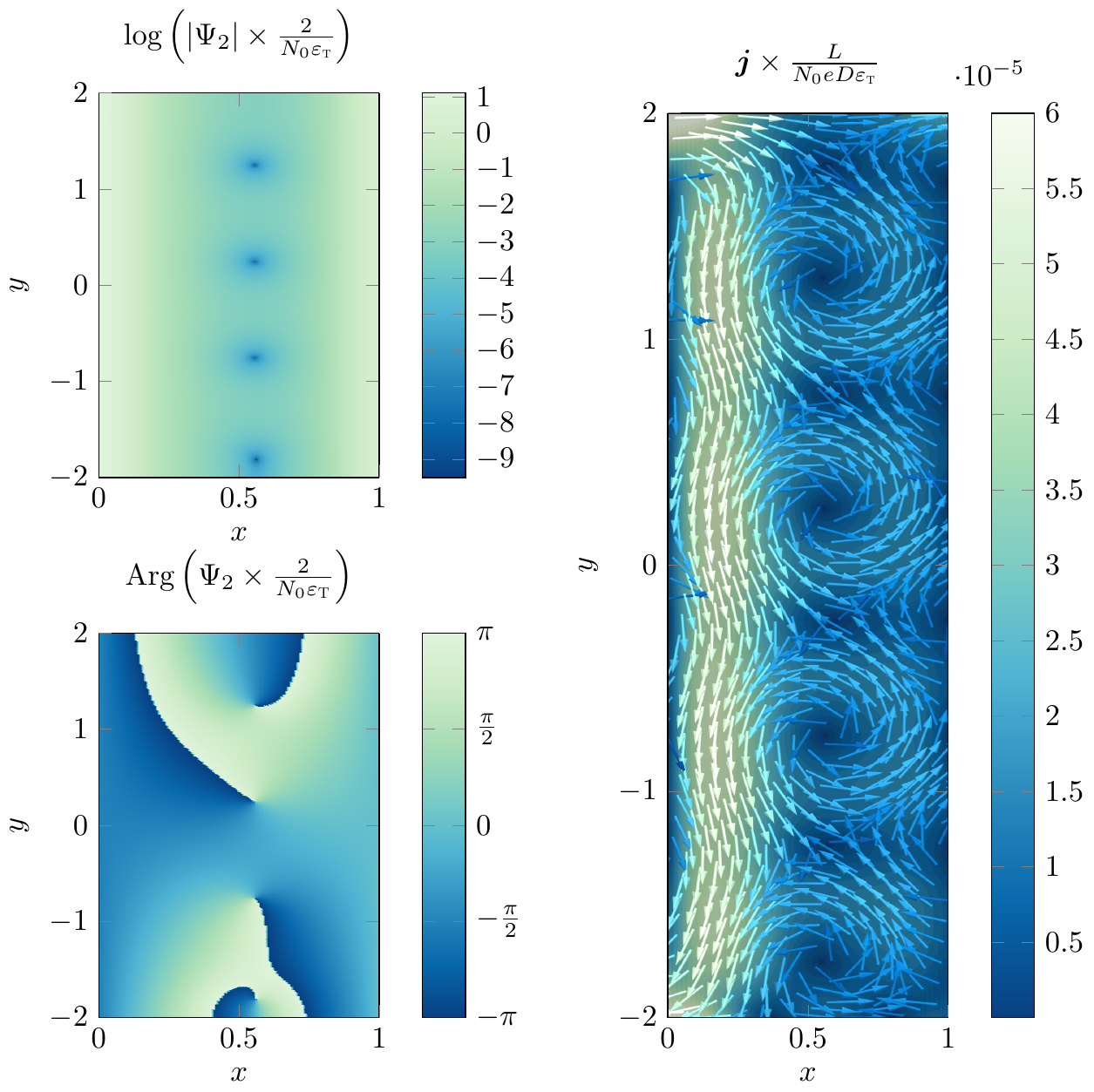}
  \caption{Amplitude and phase of the order parameter $\ordTwo$ and amplitude and direction of the fully spin-polarized supercurrent $\v j$. Here $n=4$, $\phi_r-\alpha_r=\pi/2$ and $\sin\theta_r = 0.5$.}
  \label{fig:nonSym}
\end{figure}

So far we have looked only at the case with $\phi_r = \alpha_r = 0$.
Choosing a nonzero value for $\phi_r - \alpha_r$ moves the vortices along the $y$-axis, and the locations correspond to those predicted by the analysis and given in \cref{eq:vortPosY}.
\Cref{fig:sym,fig:nonSym} shows the amplitude and phase of $\ordTwo$ as well as the supercurrent $\v j$ for a wide junction subjected to a magnetic flux of $4\Phi_0$, with $\sin\theta_r = 1$, $\phi_r - \alpha_r = 0$ and $\sin\theta_r = 0.5$, $\phi_r - \alpha_r = \pi/2$, respectively.
In \cref{fig:nonSym} the vortices are shifted $W/16$ down along the $y$-axis, which is exactly what is predicted by \cref{eq:vortPosY}.
Note also that the vortices are moved towards the side where $\sin\theta$ is smaller in \cref{fig:nonSym}.

The dependence of the vortex position on both $\alpha$ and $\theta$ suggests an experimental method to determine the effective magnetization angles describing disorder in the form of interfacial misaligned moments or artificially inserted misaligned magnetic layers in half-metallic hybrid structures.
For a fixed value of the magnetic flux and phase difference $\phi_r-\phi_l$ (which is tunable by the applied current), the $y$-coordinates of the vortices gives information about the azimuthal angles $\alpha_l$ and $\alpha_r$, while the $x$-coordinates gives information about the polar angles $\theta_l$ and $\theta_r$. This approach could possibly be easier than trying to measure the magnetization angles directly, especially if the non-collinear magnetization angle at the interface is produced by the natural misalignment of magnetic dipoles arising from the conjunction of different atomic structures at the interface.

\section{Conclusion}
Using the quasiclassical Usadel theory we have found both analytically and numerically that superconducting vortices occur also in purely odd-frequency superconducting condensates that exist in proximized half-metallic ferromagnets.
Because half-metals only have one conducting spin-band, the vortcex cores are circulated by fully polarized spin supercurrents.
An additional feature of vortex structures in half-metals compared to a normal metal is that the vortex position depend on the magnetization angles.
We suggest that this insight can be used to help determine these angles in hybrid structures.

The study of vortices in odd-frequency superconducting condensates naturally raises the question of how to define the superconducting order parameter.
Comparing the roots of the superconducting order parameter to the location vorticies we have found that the order parameter that works best is that which is made even in time by differentiation, $\frac{\partial}{\partial t} \expval{\psi_\up(\vec r, t)\psi_\up(\vec r, 0)}\rvert_{t=0}$.

\begin{acknowledgments}
  We thank M. Amundsen for helpful discussions. This work was supported by the Research Council of Norway through grant 240806, and its Centres of Excellence funding scheme grant 262633 ``\emph{QuSpin}''.
\end{acknowledgments}



\newpage
\bibliography{bibliography}

\end{document}